\documentclass[12pt,english]{article}
\usepackage[T1]{fontenc}
\usepackage[latin1]{inputenc}
\usepackage{graphicx}

\makeatletter

\usepackage{moriond}
\usepackage{axodraw}

\def\refe@jnl#1{{#1}}
\def\aj{\refe@jnl{Astron.~J.}}
\def\araa{\refe@jnl{Annu.~Rev.~Astron.~Astrophys.}}
\def\apj{\refe@jnl{Astrophys.~J.}}
\def\apjl{\refe@jnl{Astrophys.~J.~Lett.}}
\def\aap{\refe@jnl{Astron.~Astrophys.}}
\def\mnras{\refe@jnl{Mon.~Not.~R.~Astron.~Soc.}}
\def\prd{\refe@jnl{Phys.~Rev.~D}}
\def\fcp{\refe@jnl{Fund.~Cos.~Phys.}}
\def\physrep{\refe@jnl{Phys.~Rep.}}
\def\physlett{\refe@jnl{Phys.~Lett.}}

\def\mdm{{m_{\rm dm}}}

\def\be{\begin{equation}}
\def\ee{\end{equation}}
\def\bea{\begin{eqnarray}}
\def\eea{\end{eqnarray}}

\usepackage{babel}
\makeatother
\begin{document}
\vspace*{4cm}

\title{Could a $\gamma$ Line Betray the Mass of Light Dark Matter?}

\author{J. Orloff\\
LPC, Universit{\'e} Blaise Pascal, 63177 Aubi{\`e}re Cedex, France}

\maketitle\abstracts{ We~\cite{BOS} compute the pair annihilation cross
  section of light dark matter scalar particles into two photons, and
  discuss the detectability of the monochromatic line associated with these
  annihilations. }

\section{Introduction}

The need for Cold Dark Matter (DM) to describe and understand how
structures in the universe hold together, has become increasingly
pressing with the impressive amout of observational data collected
in the last ten years. Alternatives to DM, like modifications of gravity,
are being put to critical and maybe fatal test by recording maps of
gravitational lensing. Indeed, the separation recently observed in
colliding clusters between the (maybe modified) gravitational deviation
of light and the normal matter that causes it, seems very contrived
with modified gravity, and very natural if collisionless DM is the
main source of gravity. Questions about the nature of DM and its
non-gravitational detection are therefore more relevant than ever.

In this context, the precise determination by INTEGRAL/SPI~\cite{Jean:2003ci}
of the characteristics of the 511~keV line emitted in our galaxy
is intriguing. Indeed, it implies wihtout any doubt that the central
bulge of our galaxy is a strong source of positrons. Astrophysical
sources like Low Mass X-ray Binaries (LMXB) and Type 1a Supernovae
(SN1A)~\cite{Knodlseder05} cannot naturally explain why this source
is at the same time steady, extended and absent in the disk. On the
contrary, production of positrons through DM annihilation is naturally
steady and concentrated in the bulge where the DM density increases:
a fit of the needed DM density profile can even be attempted~\cite{ascasibar},
yielding a reasonable NFW profile $\rho(r)\sim r^{-1}$ for DM annihilation
at rest, in opposition to a less reasonable $\rho\sim r^{-2}$ for
DM decay.

In order to maximize the electron-positron annihilation channel, such
DM must be light (LDM)~\cite{bf,bens}, at least below muon pair threshold:
$m_{dm}<100\,\mathrm{MeV}$. More constraining upper bounds can be
obtained by a careful study of final state radiation processes ($\mbox{dm}\ \mbox{dm}^{\star}\rightarrow e^{+}\, e^{-}\,\gamma$)
and positron anihilation in flight, both producing continuous gamma
ray spectra which increase with $m_{dm}$. From the first, $m_{dm}<35$~MeV
is obtained~\cite{Boehm:2006df}, and $m_{dm}<20$~MeV from comparing the second
with error bars on the measured spectra. On the other hand, $m_{dm}$
should be higher than 2~MeV to avoid spoiling nucleosynthesis~\cite{Serpico:2004nm},
and higher than 10~MeV if there is a significant coupling to neutrinos
which can alter supernova explosions~\cite{SN}, but such is not necessarily
the case.

A fairly unique viable model satisfying all the above constraints
contains scalar DM particles with $m_{dm}\sim10\,\mathrm{MeV}$, annihilating
at rest in the galactic bulge into $e^{+}e^{-}$ pairs via $t$-channel
exchange of heavy ($>100\,\mathrm{GeV}$) fermions $F_{e}$. Given
the large local DM abundance inferred from the rotation curve, the
annihilation cross-section yielding the observed positron source is
however too small to explain a correct relic density inferred e.g.
from cosmic microwave background measurements. A further light vector
particle can then be invoked to mediate an $s$-channel annihilation
process: being velocity dependent, this process becomes dominant in
the early universe and can independently be adjusted to the relic
density.

If correct, such a model~\cite{bf} would profoundly alter the road
to unification in particle physics. It therefore seems important to
look for other experimental cross-checks. The simplest and most convincing
one would be the discovery of another gamma ray line, from the process
$\mbox{dm}\ \mbox{dm}^{\star}\rightarrow\gamma\gamma$. In the following,
we show~\cite{BOS}
 this line is anavoidable in such a model, estimate its intensity,
and discuss its observability.

\section{Dark matter annihilation cross section into two photons}

The model considered is specified by the Lagrangian ${\cal L}=\bar{\psi}_{F_{e}}(c_{r}P_{L}+c_{l}P_{R})\psi_{e}\phi_{dm}+h.c$
where $P_{R,L}$ are the chiral projectors $(1\pm\gamma_{5})/2$.
The relevant annihilation diagrams are box-diagrams containing 1,
2 or 3 heavy fermions $F_{e}$. Assuming that $dm\neq dm^{\star}$
(which fixes the circulation of arrows), there are 6 diagrams, taking
into account permutation of the 2 photon external legs.

From naive power counting, each box is logarithmically divergent.
However, gauge invariance dictates a result proportional to $F_{\mu\nu}^{2}$
rather than $A_{\mu}^{2}$. This requires 2 powers of external momenta,
so that the integrand must in fact converge like $d^{4}k/k^{6}$ for
large loop momenta $k$. In the limit $m_{F_{e}}\gg m_{e,dm}$ (relevant
due to LEP and other collider/accelerator constraints), the contribution
of momenta larger than $m_{F_{e}}$ is $\sim1/m_{F_{e}}^{2}$. The
leading $1/m_{F_{e}}$ term can thus be safely obtained by expanding
the integrand in powers of $1/m_{F_{e}}$ and keeping only the first
term. 

%\includegraphics[width=7cm]{fig3-2}$\begin{array}{c}
%\Leftrightarrow{\cal L}_{eff}=\frac{1}{m_{F_{e}}}\phi_{dm}^{*}\phi_{dm}\bar{\psi}_{e}(a+ib\gamma_{5})\psi_{e}\\
%\,\\
%\,\end{array}$
\noindent
\raisebox{-25pt}{\setlength{\unitlength}{0.7pt}\SetScale{0.7}
\begin{picture}(180,90) (60,-12)
\SetWidth{0.5}
\SetColor{Black}
\DashLine(60,0)(120,0){10}
\DashLine(120,60)(60,60){10}
\SetWidth{3}
\ArrowLine(120,0)(120,60)
\SetWidth{0.5}
\ArrowLine(120,60)(180,60)
\ArrowLine(180,0)(120,0)
\ArrowLine(180,60)(180,0)
\Photon(180,60)(240,60){3}{5}
\Photon(180,0)(240,0){3}{5}
\Text(125,30)[lc]{{\Black{$F$}}}
\Text(60,58)[lt]{{\Black{$dm$}}}
\Text(60,2)[lb]{{\Black{$dm^*$}}}
\Text(150,57)[ct]{{\Black{$e$}}}
\Text(150,3)[cb]{{\Black{$e$}}}
\Text(175,30)[rc]{{\Black{$e$}}}
\Text(120,62)[cb]{{\Black{$c^*_{R,L}$}}}
\Text(120,-2)[ct]{{\Black{$c_{L,R}$}}}
\end{picture}}
{\hspace{0.3em} {\bf +} \hspace{0.3em}} 
\raisebox{-25pt}{\setlength{\unitlength}{0.7pt}\SetScale{0.7}
\begin{picture}(180,90) (60,-12)
\SetWidth{0.5}
\SetColor{Black}
\DashLine(60,0)(150,0){10}
\DashLine(150,60)(60,60){10}
\ArrowLine(150,0)(120,30)
\ArrowLine(120,30)(150,60)
\SetWidth{3}
\ArrowLine(150,60)(180,30)
\ArrowLine(180,30)(150,0)
\SetWidth{0.5}
\Photon(180,30)(240,60){3}{5}
\Photon(120,30)(240,0){3}{7}
\Text(60,58)[lt]{{\Black{$dm$}}}
\Text(60,2)[lb]{{\Black{$dm^*$}}}
\Text(170,50)[lb]{{\Black{$F$}}}
\Text(170,10)[lt]{{\Black{$F$}}}
\Text(133,47)[rb]{{\Black{$e$}}}
\Text(133,13)[rt]{{\Black{$e$}}}
\Text(150,62)[cb]{{\Black{$c_{L,R}$}}}
\Text(150,-2)[ct]{{\Black{$c^*_{R,L}$}}}
\end{picture}
}
\hspace{0.3em}{\bf +}\hspace{0.3em}
\raisebox{-25pt}{\setlength{\unitlength}{0.7pt}\SetScale{0.7}
\begin{picture}(180,90) (60,-12)
\SetWidth{0.5}
\SetColor{Black}
\DashLine(60,0)(120,0){10}
\DashLine(120,60)(60,60){10}
\ArrowLine(120,0)(120,60)
\SetWidth{3}
\ArrowLine(120,60)(180,60)
\ArrowLine(180,0)(120,0)
\ArrowLine(180,60)(180,0)
\SetWidth{0.5}
\Photon(180,60)(240,60){3}{5}
\Photon(180,0)(240,0){3}{5}
\Text(125,30)[lc]{{\Black{$e$}}}
\Text(60,58)[lt]{{\Black{$dm$}}}
\Text(60,2)[lb]{{\Black{$dm^*$}}}
\Text(150,55)[ct]{{\Black{$F$}}}
\Text(150,5)[cb]{{\Black{$F$}}}
\Text(175,30)[rc]{{\Black{$F$}}}
\Text(120,62)[cb]{{\Black{$c_{L,R}$}}}
\Text(120,-2)[ct]{{\Black{$c^*_{R,L}$}}}
\end{picture}
}
\\[0.6em]
{$\approx$\hspace{0.3em}} 
\raisebox{-25pt}{\setlength{\unitlength}{0.7pt}\SetScale{0.7}
\begin{picture}(160,90) (60,-12)
\SetWidth{0.5}
\SetColor{Black}
\DashLine(60,0)(120,30){10}
\DashLine(120,30)(60,60){10}
\SetWidth{0.5}
\ArrowLine(120,30)(180,60)
\ArrowLine(180,60)(180,0)
\ArrowLine(180,0)(120,30)
\GOval(120,30)(5,5)(0){0.882}
\Photon(180,60)(220,60){3}{3}
\Photon(180,0)(220,0){3}{3}
\Text(60,50)[lt]{{\Black{$dm$}}}
\Text(60,10)[lb]{{\Black{$dm^*$}}}
\Text(150,57)[ct]{{\Black{$e$}}}
\Text(150,3)[cb]{{\Black{$e$}}}
\Text(175,30)[rc]{{\Black{$e$}}}
\Text(120,20)[ct]{{\Black{$\frac{a+ib\gamma_5}{m_F}$}}}
\end{picture}
}
\hspace{0.3em}{\bf +}\hspace{0.3em} $O(\frac{1}{m_F^2})
\qquad\Leftrightarrow\qquad{\cal L}_{eff}=\frac{1}{m_{F_{e}}}\phi_{dm}^{*}\phi_{dm}\bar{\psi}_{e}(a+ib\gamma_{5})\psi_{e}
$

This corresponds to {}``pinching'' the box with one $F_{e}$ into
a triangle involving only electrons and an effective dm-dm-e-e coupling
given bywith the real couplings $a,b$ given by $a+ib=c_{l}^{*}c_{r}$.
For this set-up, computing the cross-section is a loop-textbook exercise
for which we find: \begin{eqnarray*}
\sigma_{\gamma\gamma}v_{r} & = & \frac{\alpha^{2}}{(2\pi)^{3}\ m_{F_{e}}^{2}}\ \frac{m_{e}^{2}}{m_{dm}^{2}}\times\left[b^{2}|2C_{0}m_{dm}^{2}|^{2}+a^{2}|1+2C_{0}(m_{e}^{2}-m_{dm}^{2})|^{2}\right].\end{eqnarray*}
$C_{0}$ is a function of $m_{e}$ and $m_{dm}$ given by the Passarino-Veltman
scalar integral. For $m_{dm}>m_{e}$, this function develops an imaginary
part corresponding to the formation of a real $e^{+}e^{-}$ pair subsequently
annihilating into 2 photons, and giving the largest contribution for
masses above 1 MeV.

For $m_{dm}\ll m_{e}$, $C_{0}$ behaves as $[-1/(2m_{e}^{2})+m_{dm}^{2}/(3m_{e}^{4})]$,
so that both terms of the cross section behave as $m_{dm}^{2}/(m_{e}m_{F_{e}})^{2}$.
This limit is relevant to estimate the effect of heavier particles
than the electron in the loop. For example, the contribution of the
$\tau$ lepton could be significant if the corresponding couplings
$(a_{\tau},b_{\tau})$ are larger than $\approx(m_{\tau}/m_{dm})\times(a_{e},b_{e})\times(m_{F_{\tau}}/m_{F_{e}})$
(with $\mdm<m_{\tau}$), i.e. if they scale at least like usual Yukawa
couplings. Since an independent detailed analysis is required to check
whether or not such couplings can pass particle physics constraints,
we prefer giving a conservative estimate based on the electron contribution
only. The latter cannot be turned off without losing the 511 keV line
signal. It therefore constitutes a safe lower bound for assessing the detectability
of the line at $E_{\gamma}=m_{dm}$.

Within the pinch approximation, the cross-section relevant for the
origin of the 511 keV emission is: \[
\sigma_{511}v_{r}=\frac{\beta_{e}}{4\pi m_{F_{e}}^{2}}\left(a^{2}\beta_{e}^{2}+b^{2}\right)\]
 with $\beta_{e}=\sqrt{1-m_{e}^{2}/m_{dm}^{2}}$, which indeed for
$b=0$ reduces to the expression used~\cite{ascasibar} for large $m_{F_{e}}$.
After careful comparison with SPI data, Ascasibar et al.~\cite{ascasibar} found \[
\sigma_{511}v_{r}=2.6\ 10^{-30}\ \left(\frac{m_{dm}}{\mbox{MeV}}\right)^{2}\mbox{cm}^{3}/\mbox{s}.\]

The $\gamma\gamma$ annihilation cross-section is then also determined
by this measurement in terms of the ratio of annihilation branching
ratios: \begin{equation}
\eta\doteq\frac{\sigma_{\gamma\gamma}}{\sigma_{511}}=\frac{\alpha^{2}}{2\pi^{2}\,\beta_{e}}\ \frac{m_{e}^{2}}{m_{dm}^{2}}\frac{a^{2}|1+2\,(m_{e}^{2}-m_{dm}^{2})C_{0}|^{2}+b^{2}|2m_{dm}^{2}C_{0}|^{2}}{a^{2}\beta_{e}^{2}+b^{2}}\label{def_eta}\end{equation}
 As announced, this ratio cannot vanish, whatever the value of $a/b$,
so that a minimum $\gamma\gamma$ flux is guaranteed. As $m_{dm}$
approaches $m_{e}$ from above, the ratio increases like $\beta_{e}^{-3}$
for a pure scalar coupling ($b=0$) and like $\beta_{e}^{-1}$ for
an axial one ($a=0$). The ratio decreases almost linearly with the
dark mater mass for $\mdm>1$ MeV. In the table below, we give typical
values of the ratio $\eta$ for the most conservative case (i.e. $a=0$,
$\beta_{e}^{-1}$): \[
\begin{array}{|r|cccc|}
\hline m_{dm}(\textrm{MeV}): & 0.52 & 1 & 5 & 20\\
\hline \eta(a=0): & 8.8\,10^{-5} & 1.4\,10^{-5} & 3.6\,10^{-6} & 8.1\,10^{-7}\\\hline \end{array}\]
Notice that the simple guess~\cite{Kasuya:2006kj} applied to the case
of decaying DM \[
\eta_{guess}\approx\frac{\alpha^{2}m_{dm}^{2}}{2\pi^{2}m_{e}^{2}\beta_{e}^{3}}\]
increases instead of decreasing with $\mdm$. For a typical mass of
10 MeV, this guess overestimates the monochromatic flux by a factor
635 with respect to our result (Eq.\ref{def_eta}). As we will see
in the next section, such a factor is crucial to the line observability.

\section{Detectability of the monochromatic line}

A few experiments have already scanned the energy range above the
electron mass. The instruments on board of INTEGRAL for example have
been designed to survey point-like objects as well as extended sources
over an energy range between 15 keV-10 MeV. The instrument INTEGRAL/SPI
itself is a spectrometer designed to monitor the 20 keV-8 MeV range
with excellent energy resolution. Therefore a legitimate question
is whether or not the line $E_{\gamma}=m_{dm}$ could have been (or
could be) detected by the same instrument that has unveiled the 511
keV signal. This essentially depends on the ratio $\eta$ as given
above, and on the background.

\begin{figure}
\includegraphics[width=8cm]{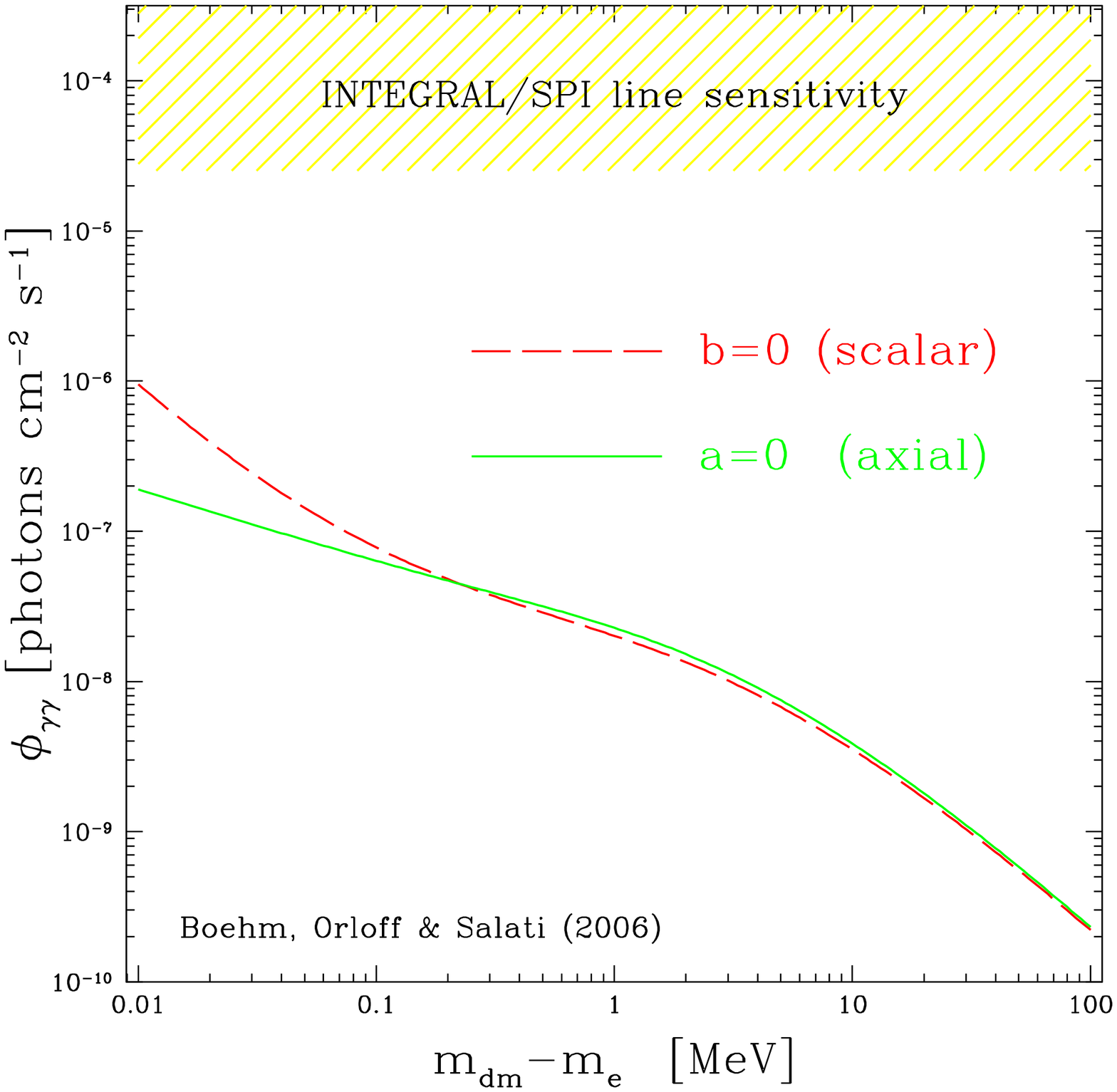}\hfill 
\includegraphics[width=8cm]{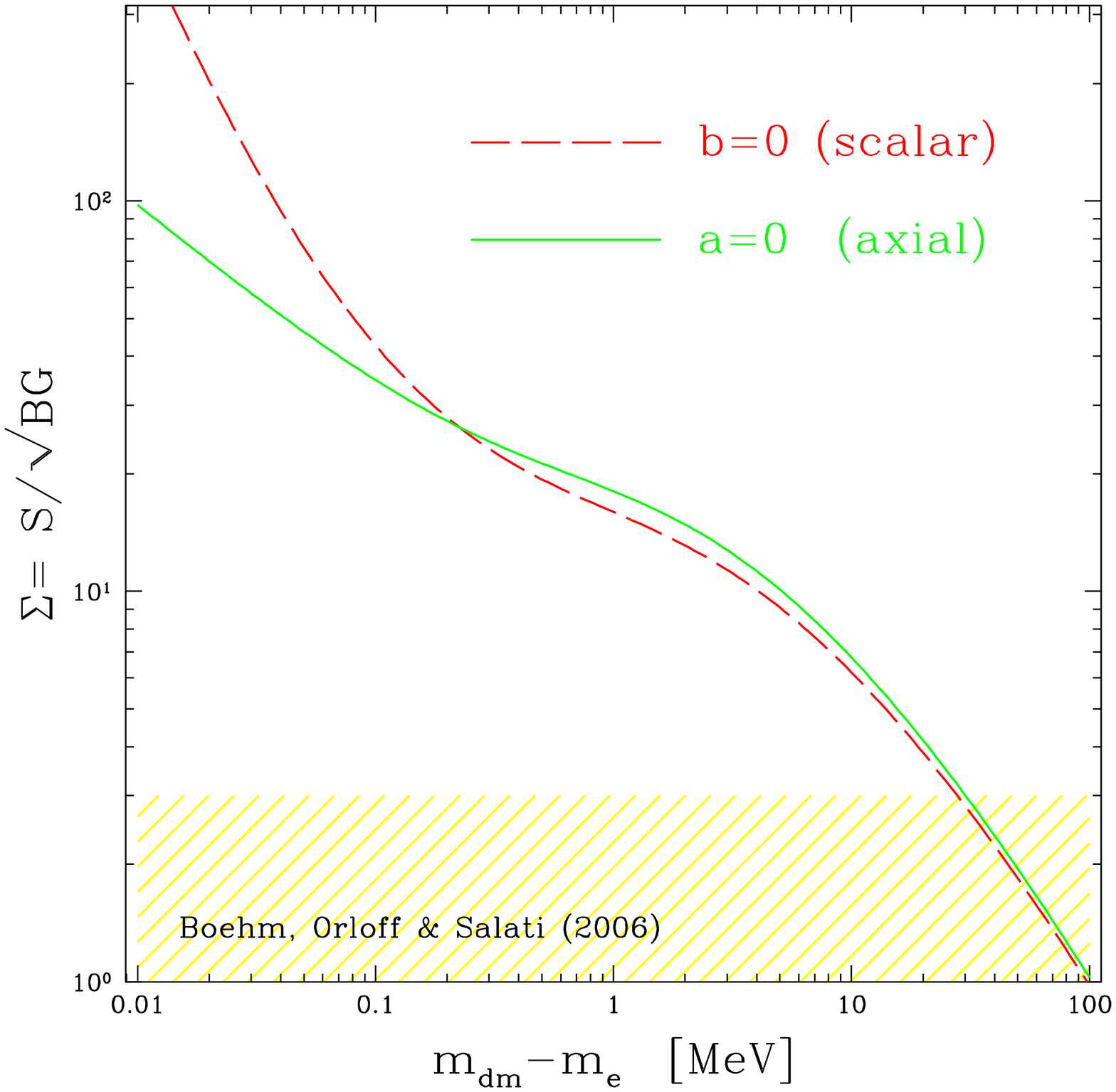}

\caption{Left: flux of the monochromatic $E_{\gamma}=m_{dm}$ line from a
8 degree cone around the galactic center. Right: significance of the
monochromatic $E_{\gamma}=m_{dm}$ line above the continuum background
for one year of observation with an ideal detector of 1 m$^{2}$ and
a $10^{-3}$ energy resolution.}

\label{sigmagg} 

\label{fluxgg} 
\end{figure}

%FFFFFFFFFFFFFFFFFFFFFFFFFFFFFFFFFFFFFFFFFFFFFFFFFFFFFFFFFFFFFFFFFFFFFFFFFFFFFF
%FFFFFFFFFFFFFFFFFFFFFFFFFFFFFFFFFFFFFFFFFFFFFFFFFFFFFFFFFFFFFFFFFFFFFFFFFFFFFF
The 511 keV emission has been measured with a $\sim$ 10\% precision
~\cite{Strong:2005zx} to be \[
\langle I_{511}\rangle=6.62\times10^{-3}\;{\rm ph\; cm^{-2}\; s^{-1}\; sr^{-1}}\]
inside a region that extends over $350^{\circ}<l<10^{\circ}$ in longitude
and $|b|<10^{\circ}$ in latitude. If this emission originates from
a NFW distribution of LDM species around the galactic center with
a characteristic halo radius of 16.7 kpc, the signal from the inner
$5^{\circ}$ is found~\cite{ascasibar} to be \[
\langle I_{511}(5^{\circ})\rangle=1.8\times10^{-2}\;{\rm ph\; cm^{-2}\; s^{-1}\; sr^{-1}}\;\;,\]
once the SPI response function is taken into account and the instrumental
background is properly modeled. If the positron propagation is negligible,
then the map of the 511 keV emission should correspond to that of
the LDM annihilations.

Within this approximation, we expect the spatial distributions of
both the 511 keV and the two-gamma ray lines to be identical and their
intensities to be related by the ratio $\eta$: \[
\langle I_{\gamma\gamma}(\theta_{\gamma\gamma})\rangle=\eta\;\frac{\theta_{511}}{\theta_{\gamma\gamma}}\;\frac{\langle I_{511}(\theta_{511})\rangle}{(1-3f/4)}\;\;.\]
 This expression is approximately valid as long as the angular radii
$\theta_{\gamma\gamma}$ and $\theta_{511}$ of the regions monitored
by the gamma-ray spectrometer are small. In what follows, the fraction
$f$ of positrons forming positronium has been taken~\cite{ascasibar}
equal to $93\%$. This is in perfect agreement with the positronium
fraction later derived~\cite{Weidenspointner:2006nu}, i.e. $f_{Ps}=0.92\pm0.09$.
The monochromatic line flux \[
\phi_{\gamma\gamma}(<\theta_{\gamma\gamma})=\pi\,\theta_{\gamma\gamma}^{2}\;\langle I_{\gamma\gamma}(\theta_{\gamma\gamma})\rangle\]
has been plotted in Fig.~\ref{fluxgg} in the case of the LDM model
with $F$ exchange and assuming a NFW profile. The angular radius
$\theta_{\gamma\gamma}=8^{\circ}$ corresponds to the field of view
of the satellite. For typical LDM masses in the MeV range, the expected
flux is about three orders of magnitude below the claimed INTEGRAL/SPI
line sensitivity~\cite{Roques:2003xg} (which is about $2.5\times10^{-5}$
ph cm$^{-2}$ s$^{-1}$ after $10^{6}$ seconds). An unrealistic exposure
of 30,000 years would thus be required in order to detect the $E=\mdm$
line. When the LDM species is degenerate in mass with the electron,
the flux is only a factor of 25 below the SPI detection limit (assuming
a pure scalar coupling b=0). As long as the mass difference $m_{dm}-m_{e}$
does not exceed 0.1 MeV, it is roughly comparable with the expected
478 keV line signal emitted by Novae~\cite{Teegarden:2006ni}, that
is about $\sim10^{-7}$ ph cm$^{-2}$ s$^{-1}$.

SPI sensitivity is limited by the instrumental background that arises
mostly from cosmic rays impinging on the apparatus and activating
the BGO scintillator. On the contrary, the absolute sensitivity of
an ideal instrument is purely limited by the gamma-ray continuum background.
This emission has been recently estimated~\cite{Strong:2005zx} \[
I_{BG}(E)=1.15\times10^{-2}\, E^{-1.82}\;{\rm ph\; cm^{-2}\; s^{-1}\; sr^{-1}\; MeV^{-1}}\;\;,\]
 inside the central region that extends over $350^{\circ}<l<10^{\circ}$
in longitude and $|b|<10^{\circ}$ in latitude. The energy $E$ is
expressed in units of MeV.

We thus estimate the significance $\Sigma\equiv\textrm{signal}/\sqrt{\textrm{background}}$
for the LDM line to emerge above this background (assuming it is isotropic)
to be \[
\Sigma=\sqrt{\pi}\;\theta_{511}\;\frac{\langle I_{511}(\theta_{511})\rangle}{(1-3f/4)}\;\eta\;\sqrt{\frac{S_{0}\, T_{0}}{I_{BG}\,\Delta E_{0}}}\;\;,\]
 with $S_{0}$ the surface of the detector, $T_{0}$ the exposure
time, $I_{BG}$ the above-mentionned continuum background intensity
and $\Delta E_{0}$ the energy resolution. The significance $\Sigma$
(displayed as a function of the dark particle mass in Fig.~\ref{sigmagg}
for a surface of 1 m$^{2}$, an exposure duration of $T_{0}=1$ year
and an energy resolution of $0.1\%$) indicates that those values
would theoretically allow to extract the minimal guaranteed signal
computed at 3 standard deviations above background for all relevant
LDM masses below 30 MeV.

There is nothing to be gained by narrowing the angular aperture $\theta_{\gamma\gamma}$
because, for the assumed NFW profile, the signal increases linearly
with this angular radius, as does the square root of an isotropic
background.

In contrast, note that the monochromatic line should be extremely
narrow: its width is expected to be about a few eV which experimentally
is very challenging if one compares it with the present SPI sensitivity
that is about $10^{-3}$ at MeV energies. At lower energies, there
are nevertheless instruments, e.g. X-ray CCD, bolometers, Bragg spectrometers
which are able to resolve eV widths. A significant improvement on
the resolution $\Delta E_{0}$ at higher energies would probably be
necessary in order to reach a large enough significance and ensure
detection. Indeed, an effective surface of 1m$^{2}$ might be hard
to attain in space.

Next generation instruments such as AGILE/(super AGILE) or GLAST,
which in principle could be more promising, will probably be limited
by the energy range that they are able to investigate. Future instruments
might nevertheless be able to see this line if their energy resolution
and sensitivity are improved by a large factor with respect to SPI
present characteristics.

Maybe a better chance to detect this line is to do observations at
a high latitude and a longitude slightly off the galactic centre.
In this case, indeed, the background should drop significantly (the
density of dark clouds has been measured recently~\cite{grenier})
but the line flux may decrease by a smaller factor.

\section*{Acknowledgement}

It is a pleasure to thank the organisers for the most enjoyable and
stimulating conference.

\bibliographystyle{unsrt}
\bibliography{line-4}

\begin{thebibliography}{10}

\bibitem{BOS}
C.~Boehm, J.~Orloff, and P.~Salati.
\newblock {\em Phys. Lett.}, B641:247--253, 2006.

\bibitem{Jean:2003ci}
P.~Jean et~al.
\newblock {\em Astron. Astrophys.}, 407:L55, 2003.

\bibitem{Knodlseder05}
J.~Knodlseder et~al.
\newblock {\em Astron. Astrophys.}, 441:513--532, 2005.

\bibitem{ascasibar}
Y.~Ascasibar, P.~Jean, C.~Boehm, and J.~Knoedlseder.
\newblock {\em Mon. Not. Roy. Astron. Soc.}, 368:1695--1705, 2006.

\bibitem{bf}
C.~Boehm and P.~Fayet.
\newblock {\em Nucl. Phys.}, B683:219--263, 2004.

\bibitem{bens}
C.~Boehm, T.~A. Ensslin, and J.~Silk.
\newblock {\em J. Phys.}, G30:279--286, 2004.

\bibitem{Boehm:2006df}
C.~Boehm and P.~Uwer.
\newblock 0600.

\bibitem{Serpico:2004nm}
Pasquale~Dario Serpico and Georg~G. Raffelt.
\newblock {\em Phys. Rev.}, D70:043526, 2004.

\bibitem{SN}
P.~Fayet, D.~Hooper, and G.~Sigl.
\newblock {\em Phys. Rev. Lett.}, 96:211302, 2006.

\bibitem{Kasuya:2006kj}
S.~Kasuya and M.~Kawasaki.
\newblock {\em Phys. Rev.}, D73:063007, 2006.

\bibitem{Strong:2005zx}
Andrew~W. Strong et~al.
\newblock {\em Astron. Astrophys.}, 444:495, 2005.

\bibitem{Weidenspointner:2006nu}
Georg Weidenspointner et~al.
\newblock 0100.

\bibitem{Roques:2003xg}
J.~P. Roques et~al.
\newblock 1000.

\bibitem{Teegarden:2006ni}
B.~J. Teegarden and K.~Watanabe.
\newblock {\em Astrophys. J.}, 646:965--981, 2006.

\bibitem{grenier}
J.-M.~Casandjian I.A.~Grenier and R.~Terrier.
\newblock {\em Science}, 307:1292, 2005.

\end{thebibliography}

\end{document}